\documentclass[twocolumn]{jpsj2} 
\title{CoO$_2$-Layer-Thickness Dependence of Magnetic Properties and Possible 
Two Different Superconducting States in Na$_x$CoO$_2 \cdot y$H$_2$O}

\author{Masahito \textsc{Mochizuki}$^{1}$\thanks{E-mail address:
mochizuki@riken.jp} and Masao \textsc{Ogata}$^{2}$}

\inst{$^1$RIKEN, Hirosawa Wako, Saitama 351-0198, Japan \\
 $^2$Department of Physics, University of Tokyo, Hongo, Bunkyo-ku, Tokyo 
113-0033, Japan}

\abst{In order to understand the experimentally proposed phase diagrams of 
Na$_x$CoO$_2 \cdot y$H$_2$O, we theoretically study the CoO$_2$-layer-thickness 
dependence of magnetic and superconducting (SC) properties 
by analyzing a multiorbital Hubbard model using the random phase approximation. 
When the Co valence $s$ is +3.4, we show that the magnetic fluctuation exhibits 
strong layer-thickness dependence where it is enhanced at finite (zero) momentum 
in the thicker (thinner) layer system. A magnetic order phase appears sandwiched 
by two SC phases, consistent with the experiments. These two SC phases have 
different pairing states where one is the singlet extended $s$-wave state and 
the other is the triplet $p$-wave state. On the other hand, only a triplet 
$p$-wave SC phase with dome-shaped behavior of $T_c$ is predicted when 
$s$=+3.5, which is also consistent with the experiments. Controversial 
experimental results on the magnetic properties are also discussed.}

\kword{Co-oxide superconductor, multiorbital Hubbard model, phase diagram}

\begin{document}
\maketitle
A possible unconventional superconductivity (SC), particularly a 
spin-fluctuation-mediated one, has been expected in 
Na$_x$CoO$_2 \cdot y$H$_2$O.~\cite{Takada03,Takada05} Hence, its normal-state magnetic 
property has been intensively studied, although the results are rather 
scattered.~\cite{Takada05,Ishida03,Fujimoto04,Kato03,Ihara05b,Ning05,Mukhamedshin05,
Higemoto04,Moyoshi06}

The bulk susceptibility $\chi$ shows Curie-Weiss type upturn below $\sim$130 K 
with decreasing temperature. 
Several groups have observed a similar upturn in Knight shift ($K$) and a linear 
$K$-$\chi$ plot in NMR and $\mu$SR measurements, suggesting an increase of 
spin fluctuation at/near $q$=0.~\cite{Takada05,Kato03,Ihara05b,Higemoto04}
Among them, Ishida $et$ $al$. studied the relation between $1/T_1T$ and 
$\chi$.~\cite{Ishida03,Ihara05b} They first reported an identical 
$T$ dependence of $1/T_1T$ and $\chi$ up to $T_c$, which suggests a dominant 
ferromagnetic (FM) fluctuation at $q$=0.~\cite{Ishida03} 
Later, they reported a slightly different behavior in another sample 
where the upturn of $\chi$ is weaker than that of $1/T_1T$, which suggests 
dominant spin fluctuations at $q\sim0$ but not at $q=0$.~\cite{Ihara05b} 
On the other hand, Ning $et$ $al.$ and Mukhamedshin $et$ $al.$ observed 
$T$-independent behavior of $K$ despite the strong $T$ dependence of 
$1/T_1T$, which suggests dominant spin fluctuations at 
$q\ne0$.~\cite{Ning05,Mukhamedshin05} 
We also note that a neutron-scattering experiment did not detect any evidence 
for spin fluctuations.~\cite{Moyoshi06}

On the other hand, a relationship between $T_c$ and CoO$_2$-layer thickness has 
been pointed out by several groups.~\cite{Lynn03,Sakurai04,Ihara04a,Ihara05a,
Sakurai05a,Sakurai05b,Michioka06,Sato06}
In particular, Sakurai $et$ $al.$ determined $x$-$T$ phase diagrams where $x$ 
is the Na content.~\cite{Sakurai05a,Sakurai05b}
Here, $x$ scales with the CoO$_2$-layer thickness, i.e., a larger-$x$ sample has 
thicker CoO$_2$ layers. 
At the same time, they found that the Co valence, $s$, is constant at $\sim$+3.4
although $x$ changes because of the presence of H$_3$O$^+$ ions in Na-layers.
As a function of $x$, they found successive three phases of SC (SC1), a magnetic
order (MO) and another SC (SC2). 
Note that $s$ is directly related to the number of $t_{2g}$-electrons per Co ion 
($n_{t2g}$) as $n_{t2g}=9-s$. 
On the other hand, for samples with a slightly different value of $s$ ($s\sim$+3.5), 
only one SC phase appears, 
and $T_c$ shows a dome-shaped behavior as a function of $x$.
This indicates that a subtle change in the lattice parameter and that in the Co valence
affect drastically the electronic properties.

\begin{figure}[tdp]
\includegraphics[scale=0.4]{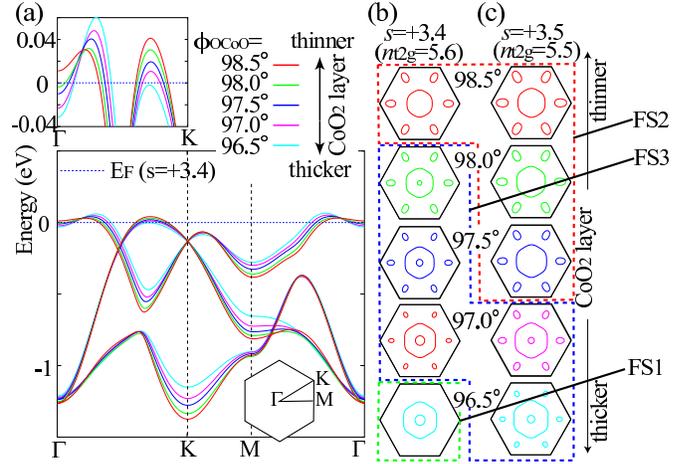}
\caption{(a) Band dispersions for various $\phi_{\rm OCoO}$ values calculated 
from the three-band tight-binding model $H_{\rm 3TB}$. 
Horizontal dashed lines denote the Fermi level for $s$=+3.4. 
(b) Deformation of the Fermi surface with varying $\phi_{\rm OCoO}$ for the 
$s$=+3.4 case. 
(c) Deformation of the Fermi surface for the $s$=+3.5 case.}
\label{Fig01}
\end{figure}

Motivated by these findings, we previously studied effects of CoO$_6$ distortion
on the band structures.~\cite{Mochizuki06}
We constructed an eleven-band tight-binding (TB) model including the Co $3d$ and the O $2p$ 
orbitals, which reproduces very well the LDA data for the bilayer-hydrate system 
of ref.~\citen{Johannes04}.
In the case of $s$=+3.4, we found that; (i) Fermi surface (FS) with double
$a_{1g}$-band cylinders around the $\Gamma$ point is realized in a system with 
thick CoO$_2$ layers. This FS was referred to as FS1 (see the bottom figure of 
Fig.~\ref{Fig01}(b)).
(ii) FS with a single $a_{1g}$-band cylinder and six ${e'}_g$-band hole pockets (FS2) 
is realized in a thin-layer system. (iii) In the moderate-thickness case, another type of 
FS with double $a_{1g}$ cylinders and six ${e'}_g$ pockets (FS3) is expected.
We discussed that this FS deformation can explain the experimental $s$=+3.4 phase 
diagram with three successive phases.~\cite{Mochizuki06} 

In this letter, we perform microscopic calculations on the SC gap structures as well as the 
nature of spin fluctuation since their knowledge is essentially important to understand 
the experimental results. We resolve the above discrepancies of the experimental results
of the character of magnetic fluctuation.
The CoO$_2$-layer-thickness dependence of magnetic fluctuation 
and SC states are studied by constructing the multiorbital Hubbard model with threefold Co 
$t_{2g}$ orbitals and by applying the random phase approximation (RPA). When $s$
is +3.4, we show that the spin fluctuation is critically enhanced with decreasing $T$
toward MO in the moderate-thickness system with FS3 (see Fig.~\ref{Fig01}(b)).
This is in agreement with the existence of MO phase in the experimental phase diagram.
By solving the Eliashberg equation, we show that the singlet 
extended $s$-wave pairing is expected in the thick-layer systems with FS1, 
while the triplet $p$-wave pairing 
is expected in the thin-layer systems with FS2. For the $s$=+3.5 case (see Fig.~\ref{Fig01}(c)), 
in contrast, the magnetic instability hardly occurs and only a triplet $p$-wave 
SC phase with dome-shaped $T_c$ behavior appears, which is also consistent with 
the experimental $s$=+3.5 phase diagram.

First, let us discuss the TB model used in this paper.
As noted before, we developed the eleven-band TB model in ref.~\citen{Mochizuki06}. But that 
model is not covenient for numerical calculations because of the large degrees of freedom.
Instead, we construct a simpler three-band TB model with only Co $t_{2g}$ orbitals, which 
reproduces the layer-thickness dependence predicted in the previous eleven-band 
analysis~\cite{Mochizuki06}.

The obtained three-band TB Hamiltonian is given by 
$H_{\rm 3TB}=\sum_{{\bf k},m,n,\sigma}\epsilon^{mn}_{\bf k}
d^{\dagger}_{{\bf k}m\sigma} d_{{\bf k}n\sigma}$ with
\begin{eqnarray}
\epsilon^{\gamma\gamma}_{\bf k}&=&\!\!
2t_1\cos k^{\gamma\gamma}_a+2t_2\cos k^{\gamma\gamma}_b+2t_2\cos
(k^{\gamma\gamma}_a+k^{\gamma\gamma}_b) \\ \nonumber
&+&\!\!2t_5\cos(2k^{\gamma\gamma}_a+k^{\gamma\gamma}_b)+2t_5\cos
(k^{\gamma\gamma}_a-k^{\gamma\gamma}_b) \\ \nonumber
&+&\!\!2t_6\cos(k^{\gamma\gamma}_a+2k^{\gamma\gamma}_b)+2t_9\cos
(2k^{\gamma\gamma}_a) \\ \nonumber
&+&\!\!2t_{10}\cos (2k^{\gamma\gamma}_b)+2t_{10}\cos
(2k^{\gamma\gamma}_a+2k^{\gamma\gamma}_b), \\
\epsilon^{\gamma\gamma'}_{\bf k}&=&\!\!
2t_3\cos k^{\gamma\gamma'}_b
+2t_4\cos (k^{\gamma\gamma'}_a+k^{\gamma\gamma'}_b)
+2t_4\cos k^{\gamma\gamma'}_a \\ \nonumber
&+&\!\!2t_7\cos (k^{\gamma\gamma'}_a+2k^{\gamma\gamma'}_b)
+2t_7\cos (k^{\gamma\gamma'}_a-k^{\gamma\gamma'}_b) \\ \nonumber
&+&\!\!2t_8\cos (2k^{\gamma\gamma'}_a+k^{\gamma\gamma'}_b)
+2t_{11}\cos (2k^{\gamma\gamma'}_b) \\ \nonumber
&+&\!\!2t_{12}\cos (2k^{\gamma\gamma'}_a+2k^{\gamma\gamma'}_b)
+2t_{12}\cos (2k^{\gamma\gamma'}_a)
-\Delta/3.
\end{eqnarray}
Here, $\gamma$ and $\gamma'$ represent $xy$, $yz$ and $zx$ orbitals, and
$\Delta$ denotes the trigonal crystal field (CF) from the O ions.
For definitions of $k^{\gamma\gamma^{(')}}_a$ and $k^{\gamma\gamma^{(')}}_b$,
one could see ref.~\citen{Mochizuki04}.

Compared with the previous TB model in ref.~\citen{Mochizuki04}, the present TB model 
is much more elaborate including additional transfer integrals. It reproduces subtle
features of LDA results of ref.~\citen{Johannes04}. 
Note that the numbering of transfer integrals slightly differs
from ref.~\citen{Mochizuki04}. 

In Fig.~\ref{Fig01}, we show (a) band dispersions and (b) FSs for 
several values of OCoO-bond angles ($\phi_{\rm OCoO}$).
Here, the angle $\phi_{\rm OCoO}$ expresses the CoO$_2$-layer thickness 
where a thinner CoO$_2$ layer has a larger $\phi_{\rm OCoO}$ value 
(see Fig.1 (a) in Ref.~\citen{Mochizuki06}).
The LDA data used in the previous study was calculated using experimental structure 
data with $\phi_{\rm OCoO}=97.5^{\circ}$.
The parameter values of the present model are 
$(t_1, t_2, ..., t_{12}, \Delta)$=(35.0, $-$22.0, 153.5, 
46.1, $-$17.7, $-$14.9, 3.10, $-$52.4, $-$41.0, $-$27.6, 8.16, 
4.98, 80.0) for the case of $\phi_{\rm OCoO}=97.5^{\circ}$ where the unit is meV.
Note that FS1 (FS2) is reproduced for the thick (thin) layer 
system, while FS3 is reproduced for the moderate case. 
This type of FS-topology variation is referred to as Case C 
in the previous study.~\cite{Mochizuki06}

By further adding the Coulomb interaction term $H_{\rm int.}$, we obtain 
the multiorbital Hubbard model; $H_{\rm mo}=H_{\rm 3TB}+H_{\rm int.}$ 
where $H_{\rm int.}=H_{U}+H_{U'}+H_{J_{\rm H}}+H_{J'}+H_{V}$. 
As in the previous studies~\cite{Mochizuki04,Yanase04},
the terms $H_{U}$ and $H_{U'}$ represent the intra- and inter-orbital Coulomb 
interactions, respectively, and $H_{J_{\rm H}}$ and $H_{J'}$ represent 
the Hund's-rule coupling and the pair hopping, respectively. 
These interactions are expressed using Kanamori parameters, 
$U$, $U'$, $J_{\rm H}$ and $J'$, which satisfy the relations; $U'=U-2J_{\rm H}$ 
and $J_{\rm H}=J'$. In this paper, we include
the last term $H_{V}=V\sum_{i,j}n_in_j$ representing the Coulomb repulsion 
between adjacent $i$ and $j$ sites.

We analyze this model by applying RPA. In the present three-orbital 
case, the Green's function $\hat{G}$ is expressed in the 3$\times$3-matrix 
form corresponding to the $xy$, $yz$ and $zx$ orbitals. The irreducible 
susceptibility $\hat{\chi}^0$ has a 9$\times$9-matrix form.
The singlet (triplet) pairing interaction $\hat{\Gamma}^{\rm s}$ 
($\hat{\Gamma}^{\rm t}$) is expressed using the interaction matrices.
For detailed expressions of $\hat{G}$, $\hat{\chi}^0$, $\hat{\Gamma}^{\rm s}$ and
$\hat{\Gamma}^{\rm t}$, see ref.~\citen{Mochizuki04}.
Note that the matrices $\hat{U}^{\rm s}(q)$ and $\hat{U}^{\rm c}(q)$
slightly differ from those in ref.~\citen{Mochizuki04} since
the present model includes long-range Coulomb repulsion $H_V$. 
The matrix elements $U^{\rm s}_{mn,\mu\nu}(q)$ ($U^{\rm c}_{mn,\mu\nu}(q)$) are 
$U$ ($U+2V({\bf q})$) for $m=n = \mu=\nu$, 
$J_{\rm H}$ ($2U'+2V({\bf q})-J_{\rm H}$) for $m=n \ne \mu=\nu$, 
$U'$ ($-U'+2J_{\rm H}$) for $m=\mu \ne n=\nu$, 
$J'$ ($J'$) for $m=\nu \ne n=\mu$, and 
0 for others, where 
$V({\bf q})=2V[\cos(q_1) + \cos(q_2) +\cos(q_1+q_2)]$ with 
$q_1=\sqrt{3}/2q_x-1/2q_y$ and $q_2=q_y$. 
We discuss the nature of SC by solving the Eliashberg equation.
Calculations are numerically carried out with 128$\times$128 $k$-meshes in the 
first Brillouin zone, and 1024 Matsubara frequencies.

\begin{figure}[tdp]
\includegraphics[scale=0.4]{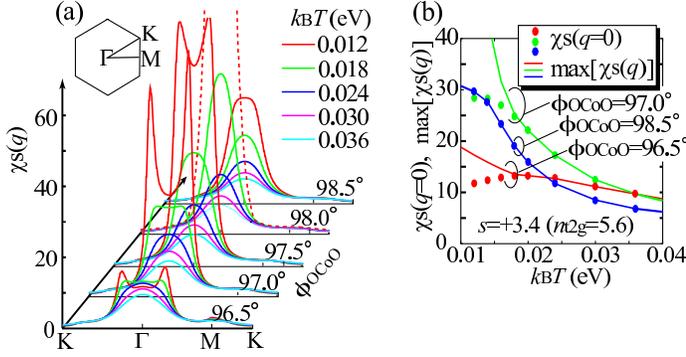}
\caption{Calculated results on the spin susceptibility $\chi_s(q)$ for 
the $s$=+3.4 case. 
(a) Momentum dependence of $\chi_s(q)$ for several $T$ and 
$\phi_{\rm OCoO}$ values. 
(b) $\chi_s(q)$ at $q$=0 ($\chi_s(q=0)$) and 
maximum $\chi_s(q)$ (max[$\chi_s(q)$])
plotted as functions of $T$ for several $\phi_{\rm OCoO}$ values.}
\label{Fig02}
\end{figure}
We first discuss the results for the $s$=+3.4 case, which are calculated 
taking $U$=0.5 eV and $J_{\rm H}$=0.05 eV. From now on, the energy unit is eV. 
In Fig.~\ref{Fig02}(a), we display the spin susceptibility $\chi_s(q)$ for several 
values of $T$ and the angle $\phi_{\rm OCoO}$, which shows strong layer-thickness 
dependence. For the thick-layer cases with FS1 ($\phi_{\rm OCoO}=96.5^{\circ}$ 
and $97.0^{\circ}$), $\chi_s(q)$ has peak structures at finite momentum $q=Q_s$ 
in the low-$T$ region. 
This $q=Q_s$ fluctuation is induced by the electron 
scattering between the inner and outer $a_{1g}$ FSs owing to the intra-orbital 
Coulomb repulsion $U$. Actually, $Q_s$ is the wave number which bridges these 
inner and outer FSs. 
On the other hand, for the thin-layer cases with FS2
($\phi_{\rm OCoO}=98.0^{\circ}$ and $98.5^{\circ}$), $\chi_s(q)$ has a FM peak at $q=0$. 
This FM fluctuation is induced by the inter-orbital Hund's-rule coupling 
$J_{\rm H}$ between $a_{1g}$ and $e'_g$ FSs as discussed previously.~\cite{Mochizuki04}

For the moderate layer-thickness systems with FS3 
($\phi_{\rm OCoO}=97.0^{\circ}$-$98.0^{\circ}$), 
a critical enhancement of $\chi_s(q)$ with decreasing $T$ occurs,
which is consistent with the existence of MO phase sandwiched by two SC
phases.~\cite{Sakurai05a,Sakurai05b}
Indeed, $\chi_s(q)$ for 
$\phi_{\rm OCoO}=98.0^{\circ}$ diverges at $k_{\rm B}T=0.012$.
This magnetic instability is caused by cooperative contributions from the 
intra-band scattering between the inner and outer $a_{1g}$ FSs and the inter-band 
scattering between the $a_{1g}$ and $e'_g$ FSs owing to the FS3 geometry. 
In addition, the structure of density of states (DOS), where both $a_{1g}$ and 
$e'_g$ orbital components are large, is also responsible for the magnetic 
instability.~\cite{Mochizuki06} 
Indeed, as will be shown later, magnetic instability does not occur 
for the $s$=+3.5 case with FS3 because of a small 
$a_{1g}$ orbital component of DOS.

The puzzle in the NMR/NQR and $\mu$SR results can be solved by considering
layer-thickness dependence of $\chi_s(q)$. 
The quantities $K$ and $1/T_1T$ scale, respectively, with $\chi_s(q=0)$ and 
maximum $\chi_s(q)$ (max[$\chi_s(q)$]). These quantities obtained in the present
theory are plotted in Fig.~\ref{Fig02}(b). 
In the thick-layer case ($\phi_{\rm OCoO}=96.5^{\circ}$) with FS1,
$\chi_s(q=0)$ is rather $T$-independent since the spin fluctuation is not FM,
which is consistent with the reports about $T$-independent NMR $K$ from 
Ning $et$ $al$ and Mukhamedshin $et$ $al$.~\cite{Ning05,Mukhamedshin05}
In the thin-layer case ($\phi_{\rm OCoO}=98.5^{\circ}$) with FS2, on the other hand, 
the spin fluctuation is FM, and $\chi_s(q=0)$ and max[$\chi_s(q)$] show an identical 
increase, which reproduces the $\chi$-($1/T_1T$) relation in ref.~\citen{Ishida03}.
Finally, in the moderately-thick case ($\phi_{\rm OCoO}=97.0^{\circ}$), 
$\chi_s(q=0)$ shows weaker increase than max[$\chi_s(q)$], which reproduces
the $\chi$-($1/T_1T$) relation in ref.~\citen{Ihara05b}.
Indeed, the sample in ref.~\citen{Ihara05b} turned out to have moderately thick 
layers according to the measured NQR frequency $\nu_Q$, 
which is consistent with the present result.

\begin{figure}[tdp]
\includegraphics[scale=0.4]{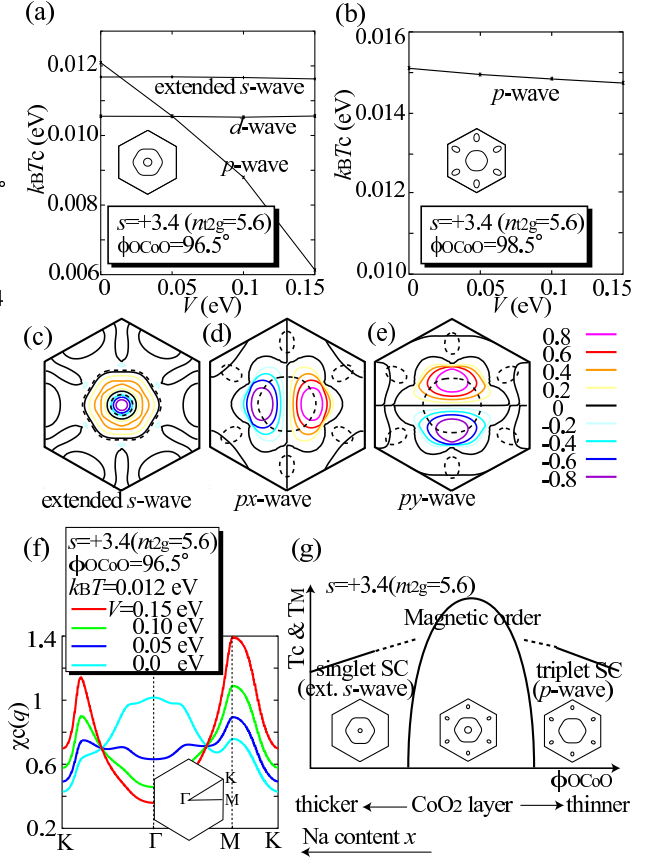}
\caption{Calculated results on superconducting properties for the $s$=+3.4 case. 
(a) $T_c$ for several pairing states plotted against $V$ for a thick-layer case with 
$\phi_{\rm OCoO}=96.5^{\circ}$. (b) that for a thin-layer case with 
$\phi_{\rm OCoO}=98.5^{\circ}$. (c) Gap structure of the extended $s$-wave pairing 
on the FS1 for $\phi_{\rm OCoO}=96.5^{\circ}$. (d) and (e) Gap structures of the 
$p_x$ and $p_y$ pairings on the FS2 for $\phi_{\rm OCoO}=98.5^{\circ}$. 
(f) Charge susceptibility $\chi_c(q)$ in the momentum space for several 
$V$ values. (g) Schematic $T$-$\phi_{\rm OCoO}$ phase diagram for the $s$=+3.4 case.}
\label{Fig03}
\end{figure}
Next, we discuss the SC properties. In Fig.~\ref{Fig03}, $T_c$ for several pairing
states are plotted as functions of $V$ both for the thick-layer case with
$\phi_{\rm OCoO}=96.5^{\circ}$ (Fig.~\ref{Fig03}(a)) and for the thin-layer case
with $\phi_{\rm OCoO}=98.5^{\circ}$ (Fig.~\ref{Fig03}(b)).
We find that in the thick-layer system with FS1, the singlet 
extended $s$-wave state is stabilized in the wide range of 
$V$ value while the triplet $p$-wave state is stabilized in the thin-layer system
with FS2. Figures~\ref{Fig03}(c)-(e) show the $k$-dependence of the obtained 
SC gaps with (c) extended $s$-wave, (d) $p_x$-wave, and (e) $p_y$-wave symmetries.
This result shows that two SC states with different symmetries are possibly
realized in this material depending on the CoO$_2$-layer thickness.

The extended $s$-wave gap here obtained for the FS1 case is equivalent to the one previously 
proposed by Kuroki $et$ $al$.~\cite{Kuroki06} The signs in this gap are the same 
within each FS but are opposite between the inner and outer FSs. This gap structure is 
stabilized not only by the spin fluctuation $\hat{\chi}^{\rm s}(q)$ but also 
by the charge fluctuation $\hat{\chi}^{\rm c}(q)$ as discussed previously. 
In fact, when $V=0$, $\lambda$ for $p$-wave state 
is slightly larger. The extended $s$-wave state dominates when we 
introduce small $V$ (at least $V/U\sim0.02$).
Note that in the expression of the singlet-pairing interaction, the 
contribution from $\hat{\chi}^{\rm s}(q)$ and that from $\hat{\chi}^{\rm c}(q)$ 
have different signs. The repulsive contribution from $\hat{\chi}^{\rm s}(q)$ 
around $q=Q_s$ favors the sign change between the inner and outer FSs. 
On the other hand, the attractive one from $\hat{\chi}^{\rm c}(q)$ favors the same 
sign in the outer FS. This is because $\hat{\chi}^{\rm c}(q)$ tends to have peak 
structures around the M-points as well as near the K-points when $V$ is 
introduced (see Fig.~\ref{Fig03}(f)). These wave numbers correspond to 
those across the outer FS. 

As for the $p$-wave state, which is realized on FS2, the $p_x$ and $p_y$ states 
are degenerate in the triangular lattice. 
Below $T_c$, a linear combination of these two 
should be realized.~\cite{Mochizuki04,Yanase04}
As shown in Figs.\ref{Fig03}(d) and (e), the gap amplitude is large on the 
$a_{1g}$ band while it is markedly small on the $e'_g$ band. 
This is in contrast to the previous results, which show dominant gaps 
on the $e'_g$ pockets.~\cite{Mochizuki04,Yanase04,Kuroki04,Mochizuki05} 
This difference is caused by the difference of TB models used. 
In the previous model, the $a_{1g}$ band has a steep slope at the Fermi level 
resulting in a small DOS, which is unfavorable for the gap opening on the $a_{1g}$ FS. 
On the other hand, the present model has the $a_{1g}$ band whose slope is rather 
gradual, resulting in a larger DOS and the dominant SC gap on the $a_{1g}$ FS. 
Since the present model reproduces the LDA band structure much more precisely,
we consider that the present result is more realistic.

Some groups observed decreasing $K$ below 
$T_c$ in the NMR experiments.~\cite{Kobayashi03b,Kobayashi05,Zheng06} 
We speculate that samples used by Kobayashi $et$ $al.$ are located in the predicted
${\it singlet}$ extended $s$-wave phase according to the measured NQR frequency 
$\nu_Q$.~\cite{SatoPC} On the other hand, $T$-independent behavior of $K$ below 
$T_c$ may be observed if we measure a sample with the triplet $p$-wave phase. 
However, synthesis of such a sample is rather difficult and should be done carefully 
since the triplet SC is quite fragile against impurities and oxygen defects. 
There exist discrepancies in the experimental data on $H_{c2}$ and 
specific heat.~\cite{Takada05} These can be also resolved if we consider the
two different SC states with different FS topologies. 
The inner $a_{1g}$ FS in FS1 and the $e'_g$ pockets 
in FS2, which reproduce the experimental data, are proved to be quite small. 
These issues will be discussed in detail elsewhere. The predicted FS deformation 
should be detected in the bilayer-hydrate materials by the bulk-sensitive angle 
resolved photoemission spectroscopy.~\cite{Shimojima06}

\begin{figure}[tdp]
\includegraphics[scale=0.4]{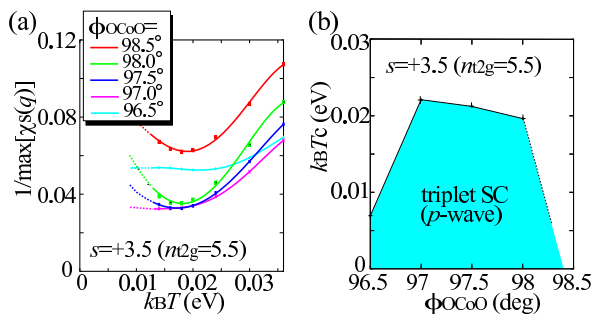}
\caption{Calculated results for the $S$=+3.5 case. (a) Inverse of maximum 
$\chi_s(q)$ plotted against $T$ for several $\phi_{\rm OCoO}$ values. 
(b) $T$-$\phi_{\rm OCoO}$ phase diagram, which shows a dome-shaped 
$T_c$ of the triplet $p$-wave pairing state.}
\label{Fig04}
\end{figure}
Finally, let us discuss briefly the case with $s$=+3.5. 
In this case, the magnetic instability is weak 
as compared to the $s$=+3.4 case. In Fig.~\ref{Fig04} (a), we show 1/max[$\chi_s(q)$] 
plotted against $T$ for several $\phi_{\rm OCoO}$ values, which are calculated 
taking $U$=0.58, $J_{\rm H}$=0.05 and $V$=0.0. This figure shows no critical 
enhancement of $\chi_s(q)$ even though we use slightly larger Coulomb parameters 
than those used in the calculations for the $s$=+3.4 case. 
As for the SC state, it is proved that only the triplet $p$-wave pairing 
is stabilized. As shown in Fig.~\ref{Fig04} (b), $T_c$ 
shows a dome-shaped behavior as a function of $\phi_{\rm OCoO}$, which is consistent 
with the experiments.~\cite{Sakurai05a,Sakurai05b}

To summarize, we have studied the CoO$_2$-layer-thickness dependence of 
magnetic and SC properties in Na$_x$CoO$_2 \cdot y$H$_2$O. By analyzing the 
multiorbital Hubbard model using RPA, we have reproduced the experimentally 
obtained $s=$+3.4 phase digram containing successive SC1, MO and SC2 phases 
as well as the $s=$+3.5 phase diagram containing one SC phase with 
dome-shaped $T_c$ behavior. We have shown that two SC phases for $s=$+3.4 
have different pairing states where one is the singlet extended $s$-wave 
state and another is the triplet $p$-wave state, while the SC phase for 
$s=$+3.5 has the $p$-wave state. We also discuss that the puzzling NMR/NQR and 
$\mu$SR results on the character of magnetic fluctuation can be understood by 
considering the strong layer-thickness dependence of the magnetic 
fluctuation.~\cite{Korshunov06}

We thank H. Sakurai, K. Ishida, Y. Yanase, K. Yoshimura, Y. Kobayashi, and M. Sato for valuable discussions. This work is supported by a Grant-in-Aid for Scientific Research from MEXT and by the RIKEN Special Postdoctoral Researcher Program.

\end{document}